\documentclass[%
 aip,
 amsmath,amssymb,
]{revtex4-1}

\usepackage{graphicx}
\usepackage{dcolumn}
\usepackage{bm}
\usepackage[utf8]{inputenc}
\usepackage[T1]{fontenc}
\usepackage{mathptmx}
\usepackage{makeidx}
\makeindex

\begin{document}

\preprint{AIP/123-QED}

\title[]{Curved surface geometry-induced topological change\\ of an excitable planar wave}

\author{Kazuya Horibe}
\email{khoribe@bio.sci.osaka-u.ac.jp}
\affiliation{ 
Department of Biological Sciences, Osaka University, Toyonaka, Osaka, 560-0043, Japan
}%
\author{Ken-ichi Hironaka}%

\affiliation{ 
Department of Biological Sciences, The University of Tokyo, Hongo, Tokyo 113-0033, Japan
}%

\author{Katsuyoshi Matsushita}
\affiliation{ 
Department of Biological Sciences, Osaka University, Toyonaka, Osaka, 560-0043, Japan
}%
\author{Koichi Fujimoto}
\affiliation{ 
Department of Biological Sciences, Osaka University, Toyonaka, Osaka, 560-0043, Japan
}%

\date{\today}

\begin{abstract}
On the curved surfaces of living and nonliving materials, planar excitable waves frequently exhibit directional change and subsequently undergo a topological change; that is, a series of wave dynamics from fusion, annihilation to splitting.
Theoretical studies have shown that excitable planar stable waves change their topology significantly depending on the initial conditions on flat surfaces, whereas the directional-change of the waves occurs based on the geometry of curved surfaces. 
However, it is not clear if the geometry of curved surfaces induces this topological change. 
In this study, we first show the curved surface geometry-induced topological changes in a planar stable wave by numerically solving an excitable reaction--diffusion equation on a bell-shaped surface. 
We determined two necessary conditions for inducing topological change: the characteristic length of the curved surface (i.e., height of the bell-shaped structure) should be larger than the width of the wave and than a threshold independent of the wave width. 
As for the geometrical mechanism of the latter, we found that a bifurcation of the globally minimum geodesics (i.e. minimal paths) on the curved surface leads to the topological change. 
These conditions imply that wave topology changes can be predicted on the basis of curved surfaces, whose structure is larger than the wave width.
\end{abstract}

\maketitle

\begin{quotation}
Biochemical, neuronal, and electrical activities often show planar excitable waves traveling on the curved surfaces of eukaryotic cells, animal brains, and hearts, respectively. 
Although the waves are used for signal processing, there is limited understanding regarding the influence of the curved surface on wave dynamics.
Through numerical simulations of a planar stable wave with excitability on a bell-shaped surface, we proved that the curved surface geometry affects the direction of the wave. 
Heavy bending of the wave leads to typical changes in the topology, namely, wave collision to splitting.
We determined the two necessary conditions for this topological change, given the stability of the wave. 
The first is that the curved structure (height) should be larger than the width of the traveling wave, whereas the second is only based on the curve geometry as follows: the ratio of the height to the width of the bell-shaped surface should be higher than a threshold.
These imply that a planar excitable wave not only changes its topology sensitively on curved structures larger than the wave width, but also robustly maintains its shape on smaller structures. 
This sensitivity and robustness are relative to the wave width and may both contribute to signal processing by changing wave topology on curved surfaces.
\end{quotation}

\begin{figure}[t]
    \hspace*{\fill}
    \includegraphics[scale=0.5]{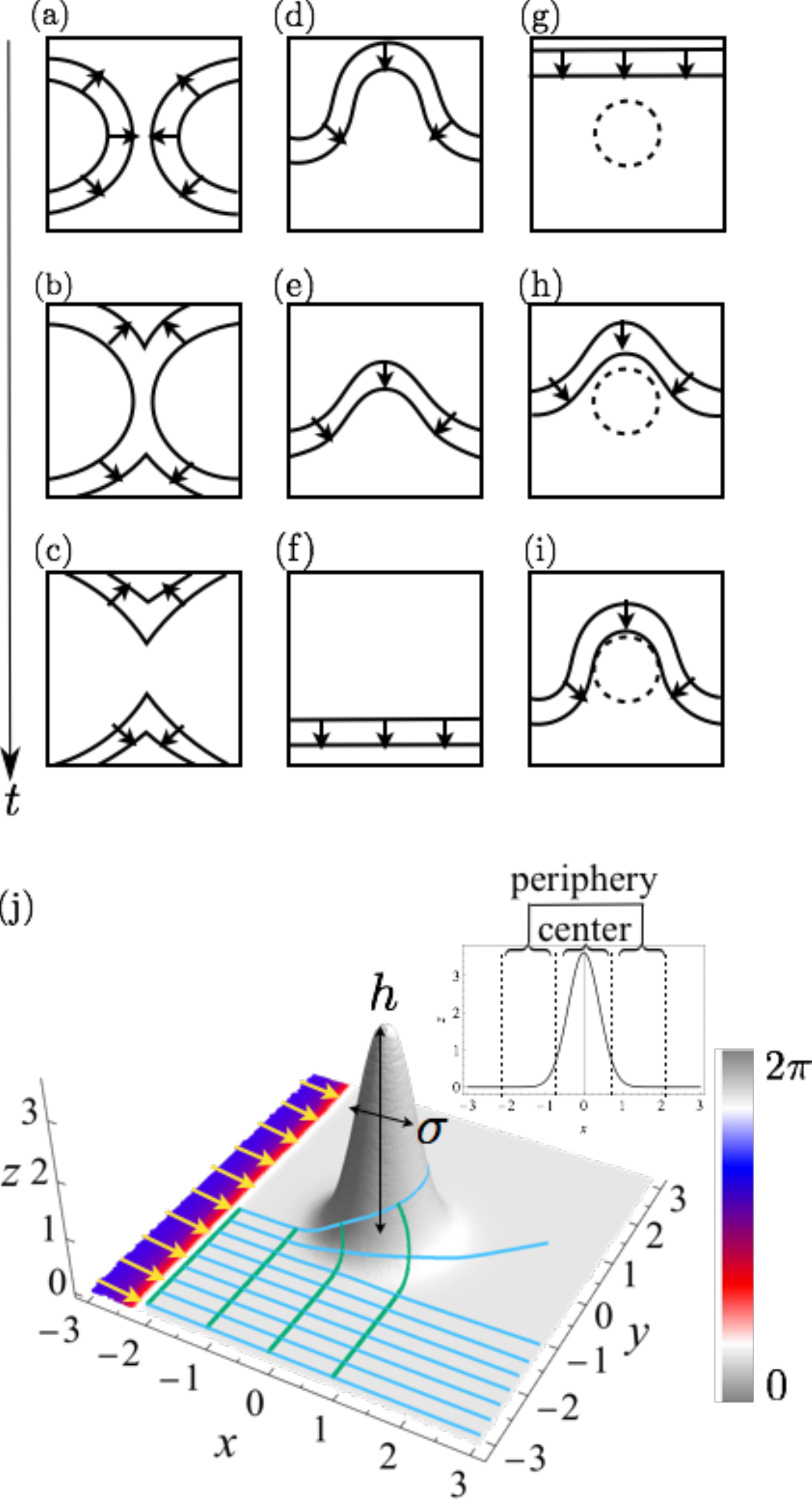}
    \hspace*{\fill}
    \caption{Conceptual scheme and model setup showing topological change of planar excitable waves.  
    (a-c) Two waves traveling in opposite directions: (a) Typical change of topology on flat surfaces, (b) collision, (c) partial annihilation, and splitting on flat surfaces. (d-f) Bent wave with stability relaxing to a planar form on a flat surface.
    (g-i) Wave bending due to a curved surface indicated by the dotted line, where the surface is higher than the typical height of a bell-shaped surface ($\sim h/2$ in (j)). 
    (j) Present model. 
    A planar wave (Eq.~\eqref{eq:HN-eq1}) initially set on the periphery of a bell-shaped surface with height $h$ and width $\sigma$ (Eq.~\eqref{eq:gaussian function}). 
    The colors indicate the wave phase (legend).
    Light blue and green lines denote the geodesics of the curved surface and the isochrones of the geodesics, respectively.
    The right upper panel is the cross section of a bell-shaped surface at $y = 0$ showing peripheral and central regions. Black (a-i) and yellow arrows (j) on waves indicate the traveling direction.}
    \label{fig:concept_setup}
\end{figure}

\section{\label{sec:level1}Introduction}
Excitable waves resulting from chemical and electrical activities appear widely on curved surfaces, including living materials such as amoeba cells~\cite{Taniguchi2013b,Gerisch2009,Gerisch2012,Gerhardt2014,Arai2010,Dobereiner2006}, animal hearts~\cite{Alonso2016,Chun2013,Neic2017} and brains~\cite{Lubenov2009,Agarwal2014,Rubino2006,Zanos2015,Martinet2017,Santos2014,Verkhlyutov2018,Roberts2019,Horning2019}. 
On curved surfaces, these traveling waves frequently exhibit a directional change and subsequently topological change, namely, a series of wave dynamics from collision, fusion, annihilation, splitting into two traveling waves~\cite{Gerhardt2014a,Alonso2016,Santos2014}. 
Because the geometry of curved surfaces is experimentally suggested to promote the directional change ~\cite{Santos2014}, thereby potentially contributing to biological signal processing ~\cite{Alonso2016,Heitmann2013,Rubino2006}, it may further promote topological change. 
However, it is unknown whether the curved surface geometry induces the topological change in excitable waves, both experimentally and theoretically.

Theoretical studies on excitable planar waves on flat surfaces, performed using reaction--diffusion (RD) systems, have shown that the abovementioned topological change is general, but it strongly depends on the initial form and arrangement of the waves~\cite{Mimura1997a,Ei2002a}. 
Two planar waves traveling in opposite directions can collide causing topological change ~(Figs.~\ref{fig:concept_setup}(a-c)), whereas a single wave with heavy bending does not change its topology when it is stable, which ensures to relax the wave to a planar form ~(Fig.~\ref{fig:concept_setup}(d-f)) ~\cite{Jones1984,Tsujikawa1989a,Biton2009a}. 
In contrast, the local surface geometry (i.e. Gaussian curvature) has been numerically shown to change the direction of excitable waves, thereby promoting the bending of the wave against the stability ~(Fig.~\ref{fig:concept_setup}(g-i))~\cite{Davydov2003,Kneer2014,Macdonald2013}.
Curved surfaces promoting the heavy bending of a wave where parts of a wave nearly collide, may lead to topological change. However, this possibility has not been examined. 

In the present study, we numerically examined whether the curved surface geometry induced topological change in a stable planar wave using an excitable RD system on a bell-shaped surface~(Fig.~\ref{fig:concept_setup}(j)). 
On a steeply curved surface, we actually showed that a planar wave bent heavily. Owing to this, it underwent the topological change. 
We determined the two conditions causing the topological change. 
As the geometrical mechanism, we found that the topological changes are accompanied with the bifurcation of the globally minimum geodesics (i.e. minimal paths) on the curved surfaces. 
These conditions imply that the wave topology change can be predicted on the curved surfaces only when the structures are larger than the wave width.

\section{\label{sec:level2}Method}
We analyzed the dynamics of an excitable planar wave on a curved surface to explore the geometry--induced topological change and the geodesics to geometrically characterize the topological change.
In this section, we introduce the RD system, curved surfaces, and their numerical methods and geodesics.

Because it is one of the simplest RD systems for representing excitable planar and stable traveling waves, we employed the FitzHugh--Nagumo equation~\cite{Nagumo1962,FitzHugh1961}:

\begin{eqnarray}
    \label{eq:HN-eq1}
\left\{
\begin{array}{l}
\frac{\partial u}{\partial t}  =  u(1-u)(u-a)-v+ D \Delta u, \\
    \frac{\partial v}{\partial t}  =  \alpha u - \beta v,
    \end{array}
\right.
\end{eqnarray}
where $u$ and $v$ are activator and inhibitor variables, respectively. 
$\Delta$ is the Laplacian. The parameter values are set as $a = 0.1$, $\alpha = 0.01$, and $\beta = -0.04$ for excitability at a stable fixed point, and $D = 0.001$ to stabilize an excitable planar traveling wave on a flat surface.

We employed a bell-shaped surface shown in Fig~\ref{fig:concept_setup}(j), which is curved in the $z$-direction on a two-dimensional $x$-$y$ plane represented by the following Gaussian function:
\begin{equation}
    \label{eq:gaussian function}
    z(x,y)=h\;\exp{(-\frac{x^2+y^2}{\sigma^2})},
\end{equation}
where $h$ and $\sigma$ are the height and width of the bell-shaped surface, respectively, used to tune the steepness of the curved surface.
The surface becomes perfectly flat at $h$ = 0 and steep with increasing $h$ or decreasing $\sigma$ at a nonzero $h$.

The simulations were performed in a rectangular region ($-L_x \leq x \leq L_x$ and $-L_y \leq y \leq L_y$, Fig.~\ref{fig:concept_setup}(j)) with an open boundary for $u$ and $v$. 
We set $L_x = L_y = 3$ except for Fig.~\ref{fig:conditions}(b) ($L_x = L_y = 1.2$ ).
In this region, we discretized the curved surface 
(Eq.~\eqref{eq:gaussian function}) 
to a triangular mesh 
through Poisson surface reconstruction ~\cite{Kazhdan2006}.
On the triangular mesh consisting of more than $10^4$ elements,
we numerically solved RD equations (Eq.~\eqref{eq:HN-eq1}) using the finite volume method~\cite{Ferziger2002} and the fourth order Runge--Kutta method with a time step $dt = 0.001$. 
As the initial condition, we set a planar traveling wave in the positive $x$-direction aligned to the boundary $x$ = -3 (Fig.~\ref{fig:concept_setup}(j)). 
Because the average element size (i.e., edge length of each triangle) was set below $1/10$ of the width of the traveling wave and elements with extremely small interior angles were manually removed, discretization errors were adequately reduced in our simulations. 

The geodesics (i.e. minimal path) of the curved surface ~(light blue lines in Fig.\ref{fig:concept_setup}(j)) were derived using the following geodesic equation~\cite{DePadua1998}:
\begin{equation}
\label{eq:geodesic equation}
\frac{d^2x^i}{dt^2}+\Gamma^i_{jk}\frac{dx^j}{dt}\frac{dx^k}{dt}=0,
\end{equation}
where $x^i$ and $\Gamma^i_{jk}$ denote the coordinates of the curved surface and the Christoffel symbols as functions of the metric $g^{ij}$, respectively:
\begin{equation}
\label{eq:Christoffel}
\Gamma^i_{jk}=\frac{1}{2}g^{il}(\frac{\partial g^{ij}}{\partial x^k}+\frac{\partial g^{lk}}{\partial x^j}-\frac{\partial g^{jk}}{\partial x^l}).
\end{equation}
We numerically solved the geodesic equation using the implicit Runge--Kutta method~\cite{Butcher1964} with a time step of $dt = 0.001$ using the NDsolve function of Mathematica (Wolfram research, USA). 
The initial position, direction, and velocity corresponded to those of the travelling wave in the simulations performed using the RD equation~(Fig.\ref{fig:concept_setup}(j)).
The initial position was set to equispaced $601$ points on $-3 \leq y \leq 3$ with $x = -3$ and $z(x,y)$ (Eq.~\eqref{eq:gaussian function}). 
The initial unit tangential vector of the geodesic equation was aligned in the positive $x$ direction.
At the same time as the simulations using Eqs.~\eqref{eq:geodesic equation} and ~\eqref{eq:Christoffel}, by mutually connecting the position on each geodesic, we obtained lines called isochrones ~(light green lines in Fig.\ref{fig:concept_setup}(j)).

\section{\label{sec:level3}Result}
\subsection{\label{sec:level3_1}Topological change in an excitable planar wave on a bell-shaped surface}
\begin{figure*}[t]
   \includegraphics[scale=0.8]{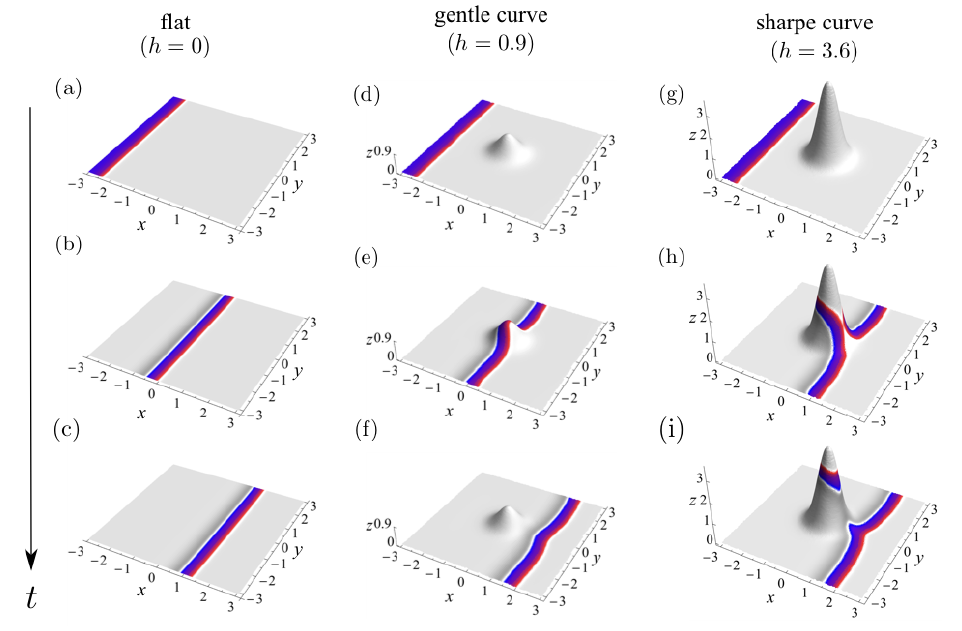}
   \caption{Presence and absence of wave topology change on a bell-shaped surface. 
   Time course of $u$ (Eq.~\eqref{eq:HN-eq1}) on a flat (a-c, (a) $t =0.5 \times 10^5$, (b) $t =2.4 \times 10^5$, (c) $t =3.5 \times 10^5$, $h=0$ in Eq.~\eqref{eq:gaussian function}), a gentle curve (d-f, (d) $t =0.5 \times 10^5$, (e) $t = 2.4 \times 10^5$, (f) $t =3.5 \times 10^5$, $h=0.9$), and a steep curve (g-i, (g) $t =0.5 \times 10^5$, (h) $t = 3.0 \times 10^5$, (i) $t =3.5 \times 10^5$, $h=3.6$). $\sigma = 0.4$. 
   The color coding is identical to that used in Fig.~\ref{fig:concept_setup}(j).
    }
    \label{fig:wave_split}
\end{figure*}

We numerically examined whether an excitable planar wave produced using the RD equation (Eq.~\eqref{eq:HN-eq1}) exhibited a topological change depending on the height $h$ of the bell-shaped surface (Eq.~\eqref{eq:gaussian function}). 
On the flat surface ($h = 0$, Figs.~\ref{fig:wave_split}(a-c)), an initially planar wave traveling from a boundary in the positive $x$-direction (Fig.~\ref{fig:wave_split}(a)) passed through the surface without changing its form, as shown in Figs.~\ref{fig:wave_split}(b) and \ref{fig:wave_split}(c).
Next, on a gently curved surface (Figs.~\ref{fig:wave_split}(d-f)), when an initially planar wave, the same as that on the flat surface, passed through (Fig.~\ref{fig:wave_split}(d)), the wave slightly bent around the bell-shaped surface (Fig.~\ref{fig:wave_split}(e)) but subsequently relaxed to the planar form, as shown in Fig.~\ref{fig:wave_split}(f).
The bending of the wave is consistent with the earlier studies on the directional change of waves due to the curved surface geometry~\cite{Davydov2003,Kneer2014,Macdonald2013}, whereas the relaxation is supported by the stability of waves (Fig~\ref{fig:concept_setup}(d-f))~\cite{Jones1984,Tsujikawa1989a}.

By contrast, on a steeply curved surface (Figs.~\ref{fig:wave_split}(g-i)), when an initially planar wave passed through  (Fig.~\ref{fig:wave_split}(g)), the wave bent very heavily around the bell-shaped surface. 
Owing to this, it subsequently exhibited the typical topology change: collision  (Fig.~\ref{fig:wave_split}(h)), partial annihilation, and splitting into two traveling waves (with circular and planar shapes, respectively, as shown in Fig.~\ref{fig:wave_split}(i)). 
Notably, this topological change is the same as that on flat surfaces, whereas the cause of the wave collision is different: heavy bending of a single wave on the curved surface and opposite traveling directions of two waves on a flat surface (Figs.~\ref{fig:concept_setup}(a-c)).
As wave collision is a precursor for this topological change and is visually distinguishable (Fig.~\ref{fig:wave_split}(h)), hereafter we identity the topological change with the wave collision.
To the best of our knowledge, this study is the first to demonstrate the topological change in a planar, stable, and excitable wave induced by a curved surface geometry.

\subsection{\label{sec:level3_2}  Conditions of curved surface geometry for wave topology change}
\begin{figure*}[t]
 \begin{center}
  \includegraphics[scale=0.6]{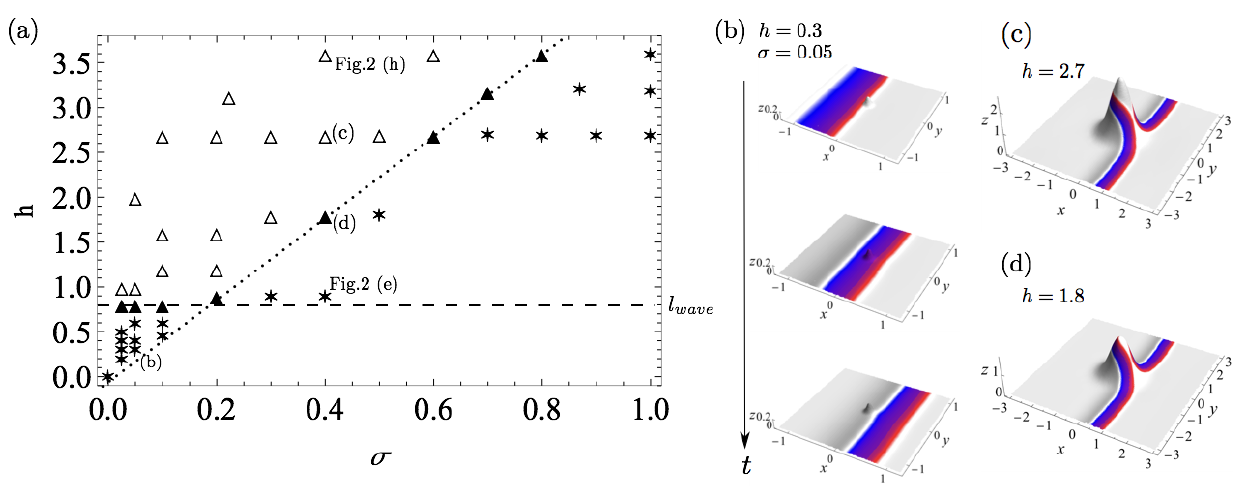}
  \caption{(a) Phase diagram of wave topology change with $h$ (height of bell-shaped surface; vertical axis) and $\sigma$ (width; horizontal axis). 
  Symbols represent presence~($\triangle$), absence~($\ast$), and threshold~($\blacktriangle$) of topological change. 
  (b)-(d) Representative examples including time evolution only at (b) parameter values indicated in (a). 
  Model parameters are (b) $h=0.3$, $\sigma = 0.05$ ,$t =0.8 \times 10^5, 1.1 \times 10^5, 1.4 \times 10^5$, and (c-d) $\sigma = 0.4$.
  The color coding for the waves is the same as that used in Fig.\ref{fig:concept_setup}(j).}
  \label{fig:conditions}
 \end{center}
\end{figure*}

We determined the presence or absence of wave topology change based on this wave collision (Fig.~\ref{fig:wave_split}(h)) in the parameter space of the bell-shaped surface, i.e., the width $\sigma$ and height $h$.
When $h$ was smaller than a threshold that was close to the width of the wave ($ l_{wave}\sim 0.8$, the dashed line in Fig.~\ref{fig:conditions}(a)), no topological change, including collision and splitting, occurred.
A traveling wave completely covered the bell-shaped surface from the center to the periphery during its passage, and it maintained its planar shape, supported by the wave stability even after the passage ~(Fig.~\ref{fig:conditions}(b)). 

By contrast, when $h$ was greater than the wave width, 
topology changes depended only on the geometrical parameters (the dotted line in Fig.~\ref{fig:conditions}).
As $h / \sigma$ decreased toward a threshold ($\sim 4.5$), the position of wave collision shifted from the periphery of bell-shaped surface to the center (Fig.~\ref{fig:conditions}(c), \ref{fig:conditions}(d) ), whereas the topological change did not occur below the threshold (Fig.~\ref{fig:conditions}(a)). 
Considering these two results, the wave maintains its planar shape when the curved structure of the surface (i.e., height) is smaller than the wave width ~($h \lesssim l_{wave}$) or less steeper~($h / \sigma \lesssim 4.5$), whereas the wave topology changes on a curved structure, which is both larger than the wave width~($h \gtrsim l_{wave}$) and steeper~($h / \sigma \gtrsim 4.5$).

\subsection{\label{sec:level3_3}The excitable planar wave on curved surfaces follows the isochrone of geodesics before the topological change
}
\begin{figure}[t]
    \includegraphics[scale=0.8]{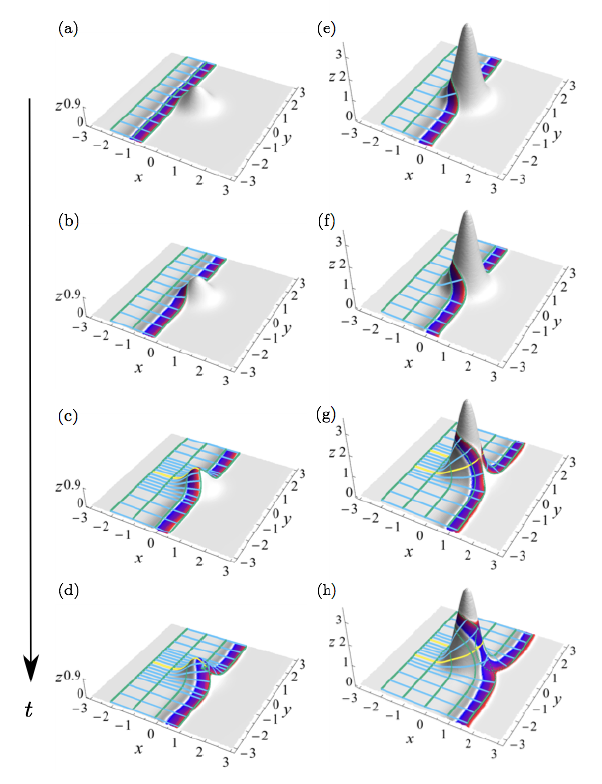}
    \caption{
    Time course of geodesic equation (Eqs.~\eqref{eq:geodesic equation} and ~\eqref{eq:Christoffel}, light blue line) from each initial condition. 
    The isochrone (light green line) in each time point ($t =1.9 \times 10^4$ (a),  $2.1 \times 10^4$ (b),  $2.3 \times 10^4$ (c), $t =2.7 \times 10^4$ (d),  $2.2 \times 10^4$ (e),  $2.6 \times 10^4$ (f), $3.0 \times 10^4$(g) and  $3.3 \times 10^4$ (h)  ). 
    Yellow lines denote the global minimal geodesic. 
    Parameter values of (a-d) and (e-h) are identical to those in Fig.\ref{fig:wave_split}(d-f) and \ref{fig:wave_split}(g-i), respectively. The color code for RD waves is the same as that in Fig.\ref{fig:concept_setup}(j).
    } 
    \label{fig:geodesic}
\end{figure}

To geometrically characterize the latter condition ($h / \sigma \gtrsim 4.5$),
we compared the spatiotemporal evolution of the wave form with the isochrones of the geodesics. 
Because excitable waves travel at a constant velocity on flat surfaces ~\cite{Tsujikawa1989a}, we first examined if the velocity was constant on the bell-shaped surface as well.
The isochrones of geodesics sufficiently agreed with the shape of the traveling wave at each time point (light green lines in Fig.\ref{fig:geodesic}(a-c),(e-g)), confirming that the wave velocity was constant even on a curved surface and the directional change followed the geodesics (light blue lines in Fig.\ref{fig:geodesic}). 
This result is consistent with a recent numerical comparison of excitable traveling spots with the geodesics of a curved surface ~\cite{Martin2018}.
This results were consistent for all regions except for those where
the wave relaxed from a V-shape owing to the stability in the absence of topological change ~(around yellow line in Fig.\ref{fig:geodesic}(d), $h / \sigma < 4.5$), and the wave collided leading to the topological change in the presence~(around yellow lines in Fig.\ref{fig:geodesic}(h), $h / \sigma > 4.5$).
In these regions, a pair of geodesics intersects $y=0$, suggesting their potential role as indicators of topological change.

The pair of geodesics intersects in the wave collision region (yellow lines in Fig.\ref{fig:geodesic}(g)) before the other geodesics intersect $y=0$. 
This indicates that the geodesic distance from the initial position (defined as $y = \pm y_c \neq 0$) to the wave collision position is shorter than that from the initial position ($y = 0$) to the center of bell shape, given the constant wave velocity on the curved surface.   
By contrast, in the absence of a topological change, the geodesics intersect near the center earlier than at the periphery (yellow line in Fig.\ref{fig:geodesic}(c,d); see also Fig.\ref{fig:concept_setup}(j)).
That is, the geodesic distance from the initial position ($y = 0$) to the center was rather shorter than that from the initial position ($y \neq  0$) to the intersection at $y=0$.
Therefore, the intersection of geodesics indicates not only the wave collision position, but also the difference on the distance between the presence and the absence of topological change. 

\begin{figure}[]
    \hspace*{\fill}
    \includegraphics[scale=0.25]{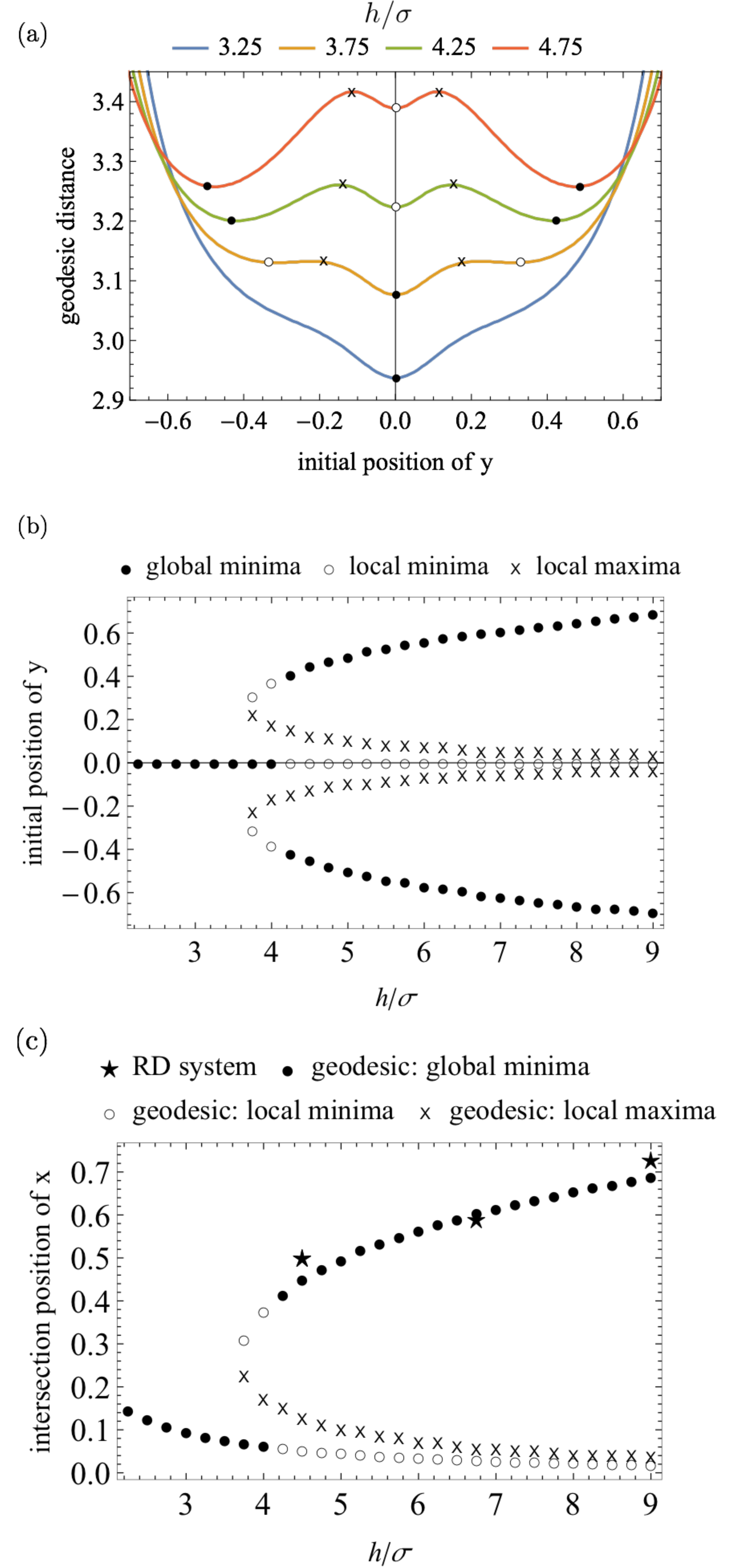}
    \hspace*{\fill}
    \caption{Bifurcation of globally minimum geodesics at the emergence of wave topology change. 
    (a) Geodesic distance to $y=0$ as a function of the initial position plotted for $-1 < y < 1$. 
    Because the geodesic starting at $y=0$ is always identical with $y=0$, the distance to $y=0$ is indefinable and is instead interpolated using the ListLinePlot function of Mathematica. The parameters are $h=1.3$ (blue line), $h=1.5$ (yellow), $h=1.7$ (green), and $h=1.9$ (red) under $\sigma=0.4$. 
    The initial position, taking the global ($\bullet$) and the local minima ($\circ$), and the local maxima ($\times$) of the geodesic distance were detected using the FindPeaks function of Mathematica. 
    (b-c) The geodesic distance plotted for the initial (b) and intersecting positions (c, average value of y-coordinate for the initial conditions with the negative and positive x-values) as a function of $h/\sigma$.  
    The definition of symbols is the same as (b), except for the wave collision position in RD system simulations ($\bigstar$). 
    }
    \label{fig:bifurcation}
\end{figure}

\subsection{\label{sec:level3_4}Bifurcation of globally minimum geodesics of curved surfaces causing the wave topology change}

We quantitatively evaluated the difference by measuring the geodesic distance from each initial position to the intersection position (Fig.\ref{fig:bifurcation}(a)). 
In the absence of a topological change, the distance mostly monotonically increased as the distance of the initial position from the x-axis increased (blue and yellow lines in Fig.\ref{fig:bifurcation}(a)), indicating that the geodesic along the x-axis is the global minimum~(yellow line in Fig.\ref{fig:geodesic}(c)).
By contrast, in the presence of a topological change, the geodesics became the globally minimum distance at the initial position of $y= \pm y_c$(green and red lines in Fig.\ref{fig:bifurcation}(a)) 
where they intersected at the wave collision position (yellow lines in Fig.\ref{fig:geodesic}(g)). 
Therefore, the globally minimum geodesics predict the wave collision position above the threshold and transition between the absence and presence of the topological change. 
This transition of globally minimum geodesics summarizes the difference of geodesics (yellow line in Fig.\ref{fig:geodesic}(c) and (g); Fig.\ref{fig:bifurcation}(a)).

To clarify the bifurcation resulting in the transition, we analyzed the behaviors of the global minima of geodesic distances and the local minima and maxima depending on $h / \sigma$ (Fig.\ref{fig:bifurcation}(a)). 
As $h / \sigma$ decreased, the initial positions taking the global minimum (i.e. $y = \pm y_c$) decreased continuously (green and red in Fig.\ref{fig:bifurcation}(a))) and switched discontinuously from $y = \pm y_c \neq 0$ to $y = 0$ (green and yellow lines in Fig.\ref{fig:bifurcation}(a)).
This is because the branches of non-zero initial positions transitioned from the global minima to the local minima ($h/\sigma \sim 4.25$ in Fig.\ref{fig:bifurcation}(b)) near the the threshold of topological change ($h/\sigma \sim 4.5$) and subsequently annihilated with the branches of the local maxima showing the saddle-node bifurcation ($h/\sigma \sim 3.5$ in Fig.\ref{fig:bifurcation}(b)). 
As long as the wave topology changes, the intersection position of geodesics, as predicted, agrees with the wave collision position in the RD system at the global minimum(Fig.\ref{fig:bifurcation}(c)). 
Therefore, the saddle-node bifurcation on the geodesic distance leads to the wave topology change.

\section{\label{sec:level4}Discussion}
\subsection{
Summary of results}
In this study, the typical wave topology change was induced by a bell-shaped surfaces~(Fig.\ref{fig:wave_split}(g-i)).
We identified two conditions that govern this topological change based on the curved surface geometry and the wave width ~(Fig.\ref{fig:conditions}). 
The first condition is that the curved structure (height of Gaussian function) should be larger than the width of the traveling wave. 
Otherwise, the stability of the travelling wave~\cite{Tsujikawa1989a} ensures that its planar shape is maintained on curved surface structures smaller than the wave width. 
The second condition is only based on the geometrical parameters, i.e., the ratio of the height $h$ to the width $\sigma$ of the Gaussian function. 
On a steeply curved surface, above the threshold of $h/\sigma$, the geodesics predicted the spatiotemporal evolution of a planar wave; therefore, the position of wave collision lead to the topology change where the globally shortest geodesics intersect each other~(Fig.\ref{fig:geodesic}).
In addition, at the threshold, a discontinuous switching of the global minimum geodesics due to a saddle-node bifurcation causes the topological change~(Fig.\ref{fig:bifurcation}). 
Thus, the geodesics provide two characteristic mechanisms regarding the second condition: (1) mutual intersection of geodesics at the wave collision position~(Fig.\ref{fig:geodesic}) and (2) the bifurcation of the globally minimum geodesics near the threshold of topological change~(Fig.\ref{fig:bifurcation}).
These conditions and mechanisms indicate that the wave topology can be adequately changed by designing the curved surfaces, instead of designing the rigid initial condition on a flat surface~(Fig.~\ref{fig:concept_setup}(a-c)).

\subsection{
Model limitation and future problems}

As this wave topology change widely appears on a flat surface, irrespective of the model details, given the stability of excitable waves, the curved-surface-geometry-induced topological change could also widely appear. 
Future studies regarding the two characteristic mechanisms of geodesics should clarify the limits and applicability to other differentiable surfaces; asymmetric Gaussian functions (e.g. different widths $\sigma$ in x- and y-axis), multiple peaks, multivalent functions such as overhanging and/or closed surfaces, and so on, which are typically observed on living and nonliving materials, should be studied.
For a planar wave losing its stability and spontaneously splitting without wave collision even on flat surfaces (e.g., in an exothermic equation with excitability) ~\cite{Mimura1997a,Ei2002a}, it would be interesting to examine the effect of wave instability on the curved-surface-geometry-induced topological change.

\subsection{
Potential functional significance}
Given the wave stability, the present geometrical mechanism is such that a planar excitable wave not only changes its topology sensitively on structures larger than the wave width on a curved surface, but it also maintains its shape robustly on smaller structures.
Considering that the wave width is approximately $5 \mu m$ in eukaryotic cells ~\cite{Taniguchi2013b} and $2 mm$ in animal brains ~\cite{Santos2014}, 
a planar stable wave robustly maintains its planar shape on surface structures (e.g., on the order of sub $\mu m$ on cells and sub $mm$
on brains) smaller than the wave width, whereas it can undergo heavy bending on larger structures, thereby changing its topology. 
Such robustness and sensitivity toward curved structures relative to wave width may contribute to the control of wave topology on living and nonliving material surfaces.

\section{\label{sec:level6}Acknowledgments}
The authors appreciate valuable comments from Dr. Gentaro Taga at the University of Tokyo. This work was partially supported by JSPS KAKENHI (Grant Number 26220004, 17H05619, 17H06386) to KF and (19K03770) to KM.

%


\begin{thebibliography}{35}%
\makeatletter
\providecommand \@ifxundefined [1]{%
 \@ifx{#1\undefined}
}%
\providecommand \@ifnum [1]{%
 \ifnum #1\expandafter \@firstoftwo
 \else \expandafter \@secondoftwo
 \fi
}%
\providecommand \@ifx [1]{%
 \ifx #1\expandafter \@firstoftwo
 \else \expandafter \@secondoftwo
 \fi
}%
\providecommand \natexlab [1]{#1}%
\providecommand \enquote  [1]{``#1''}%
\providecommand \bibnamefont  [1]{#1}%
\providecommand \bibfnamefont [1]{#1}%
\providecommand \citenamefont [1]{#1}%
\providecommand \href@noop [0]{\@secondoftwo}%
\providecommand \href [0]{\begingroup \@sanitize@url \@href}%
\providecommand \@href[1]{\@@startlink{#1}\@@href}%
\providecommand \@@href[1]{\endgroup#1\@@endlink}%
\providecommand \@sanitize@url [0]{\catcode `\\12\catcode `\$12\catcode
  `\&12\catcode `\#12\catcode `\^12\catcode `\_12\catcode `\%12\relax}%
\providecommand \@@startlink[1]{}%
\providecommand \@@endlink[0]{}%
\providecommand \url  [0]{\begingroup\@sanitize@url \@url }%
\providecommand \@url [1]{\endgroup\@href {#1}{\urlprefix }}%
\providecommand \urlprefix  [0]{URL }%
\providecommand \Eprint [0]{\href }%
\providecommand \doibase [0]{http://dx.doi.org/}%
\providecommand \selectlanguage [0]{\@gobble}%
\providecommand \bibinfo  [0]{\@secondoftwo}%
\providecommand \bibfield  [0]{\@secondoftwo}%
\providecommand \translation [1]{[#1]}%
\providecommand \BibitemOpen [0]{}%
\providecommand \bibitemStop [0]{}%
\providecommand \bibitemNoStop [0]{.\EOS\space}%
\providecommand \EOS [0]{\spacefactor3000\relax}%
\providecommand \BibitemShut  [1]{\csname bibitem#1\endcsname}%
\let\auto@bib@innerbib\@empty
\bibitem [{\citenamefont {Taniguchi}\ \emph {et~al.}(2013)\citenamefont
  {Taniguchi}, \citenamefont {Ishihara}, \citenamefont {Oonuki}, \citenamefont
  {Honda-Kitahara}, \citenamefont {Kaneko},\ and\ \citenamefont
  {Sawai}}]{Taniguchi2013b}%
  \BibitemOpen
  \bibfield  {author} {\bibinfo {author} {\bibfnamefont {D.}~\bibnamefont
  {Taniguchi}}, \bibinfo {author} {\bibfnamefont {S.}~\bibnamefont {Ishihara}},
  \bibinfo {author} {\bibfnamefont {T.}~\bibnamefont {Oonuki}}, \bibinfo
  {author} {\bibfnamefont {M.}~\bibnamefont {Honda-Kitahara}}, \bibinfo
  {author} {\bibfnamefont {K.}~\bibnamefont {Kaneko}}, \ and\ \bibinfo {author}
  {\bibfnamefont {S.}~\bibnamefont {Sawai}},\ }\bibfield  {title} {\enquote
  {\bibinfo {title} {{Phase geometries of two-dimensional excitable waves
  govern self-organized morphodynamics of amoeboid cells}},}\ }\href {\doibase
  10.1073/pnas.1218025110} {\bibfield  {journal} {\bibinfo  {journal}
  {Proceedings of the National Academy of Sciences}\ }\textbf {\bibinfo
  {volume} {110}},\ \bibinfo {pages} {5016--5021} (\bibinfo {year}
  {2013})}\BibitemShut {NoStop}%
\bibitem [{\citenamefont {Gerisch}\ \emph {et~al.}(2009)\citenamefont
  {Gerisch}, \citenamefont {Ecke}, \citenamefont {Schroth-Diez}, \citenamefont
  {Gerwig}, \citenamefont {Engel}, \citenamefont {Maddera},\ and\ \citenamefont
  {Clarke}}]{Gerisch2009}%
  \BibitemOpen
  \bibfield  {author} {\bibinfo {author} {\bibfnamefont {G.}~\bibnamefont
  {Gerisch}}, \bibinfo {author} {\bibfnamefont {M.}~\bibnamefont {Ecke}},
  \bibinfo {author} {\bibfnamefont {B.}~\bibnamefont {Schroth-Diez}}, \bibinfo
  {author} {\bibfnamefont {S.}~\bibnamefont {Gerwig}}, \bibinfo {author}
  {\bibfnamefont {U.}~\bibnamefont {Engel}}, \bibinfo {author} {\bibfnamefont
  {L.}~\bibnamefont {Maddera}}, \ and\ \bibinfo {author} {\bibfnamefont
  {M.}~\bibnamefont {Clarke}},\ }\bibfield  {title} {\enquote {\bibinfo {title}
  {{Self-organizing actin waves as planar phagocytic cup structures}},}\ }\href
  {\doibase 10.4161/cam.3.4.9708} {\bibfield  {journal} {\bibinfo  {journal}
  {Cell Adhesion and Migration}\ }\textbf {\bibinfo {volume} {3}},\ \bibinfo
  {pages} {373--382} (\bibinfo {year} {2009})}\BibitemShut {NoStop}%
\bibitem [{\citenamefont {Gerisch}\ \emph {et~al.}(2012)\citenamefont
  {Gerisch}, \citenamefont {Schroth-Diez}, \citenamefont
  {M{\"{u}}ller-Taubenberger},\ and\ \citenamefont {Ecke}}]{Gerisch2012}%
  \BibitemOpen
  \bibfield  {author} {\bibinfo {author} {\bibfnamefont {G.}~\bibnamefont
  {Gerisch}}, \bibinfo {author} {\bibfnamefont {B.}~\bibnamefont
  {Schroth-Diez}}, \bibinfo {author} {\bibfnamefont {A.}~\bibnamefont
  {M{\"{u}}ller-Taubenberger}}, \ and\ \bibinfo {author} {\bibfnamefont
  {M.}~\bibnamefont {Ecke}},\ }\bibfield  {title} {\enquote {\bibinfo {title}
  {{PIP3 waves and PTEN dynamics in the emergence of cell polarity}},}\ }\href
  {\doibase 10.1016/j.bpj.2012.08.004} {\bibfield  {journal} {\bibinfo
  {journal} {Biophysical Journal}\ }\textbf {\bibinfo {volume} {103}},\
  \bibinfo {pages} {1170--1178} (\bibinfo {year} {2012})}\BibitemShut {NoStop}%
\bibitem [{\citenamefont {Gerhardt}\ \emph
  {et~al.}(2014{\natexlab{a}})\citenamefont {Gerhardt}, \citenamefont {Ecke},
  \citenamefont {Walz}, \citenamefont {Stengl}, \citenamefont {Beta},\ and\
  \citenamefont {Gerisch}}]{Gerhardt2014}%
  \BibitemOpen
  \bibfield  {author} {\bibinfo {author} {\bibfnamefont {M.}~\bibnamefont
  {Gerhardt}}, \bibinfo {author} {\bibfnamefont {M.}~\bibnamefont {Ecke}},
  \bibinfo {author} {\bibfnamefont {M.}~\bibnamefont {Walz}}, \bibinfo {author}
  {\bibfnamefont {A.}~\bibnamefont {Stengl}}, \bibinfo {author} {\bibfnamefont
  {C.}~\bibnamefont {Beta}}, \ and\ \bibinfo {author} {\bibfnamefont
  {G.}~\bibnamefont {Gerisch}},\ }\bibfield  {title} {\enquote {\bibinfo
  {title} {{Actin and PIP3 waves in giant cells reveal the inherent length
  scale of an excited state}},}\ }\href {\doibase 10.1242/jcs.156000}
  {\bibfield  {journal} {\bibinfo  {journal} {Journal of Cell Science}\
  }\textbf {\bibinfo {volume} {127}},\ \bibinfo {pages} {4507--4517} (\bibinfo
  {year} {2014}{\natexlab{a}})}\BibitemShut {NoStop}%
\bibitem [{\citenamefont {Arai}\ \emph {et~al.}(2010)\citenamefont {Arai},
  \citenamefont {Shibata}, \citenamefont {Matsuoka}, \citenamefont {Sato},
  \citenamefont {Yanagida},\ and\ \citenamefont {Ueda}}]{Arai2010}%
  \BibitemOpen
  \bibfield  {author} {\bibinfo {author} {\bibfnamefont {Y.}~\bibnamefont
  {Arai}}, \bibinfo {author} {\bibfnamefont {T.}~\bibnamefont {Shibata}},
  \bibinfo {author} {\bibfnamefont {S.}~\bibnamefont {Matsuoka}}, \bibinfo
  {author} {\bibfnamefont {M.~J.}\ \bibnamefont {Sato}}, \bibinfo {author}
  {\bibfnamefont {T.}~\bibnamefont {Yanagida}}, \ and\ \bibinfo {author}
  {\bibfnamefont {M.}~\bibnamefont {Ueda}},\ }\bibfield  {title} {\enquote
  {\bibinfo {title} {{Self-organization of the phosphatidylinositol lipids
  signaling system for random cell migration}},}\ }\href {\doibase
  10.1073/pnas.0908278107} {\bibfield  {journal} {\bibinfo  {journal}
  {Proceedings of the National Academy of Sciences}\ }\textbf {\bibinfo
  {volume} {107}},\ \bibinfo {pages} {12399--12404} (\bibinfo {year}
  {2010})}\BibitemShut {NoStop}%
\bibitem [{\citenamefont {D{\"{o}}bereiner}\ \emph {et~al.}(2006)\citenamefont
  {D{\"{o}}bereiner}, \citenamefont {Dubin-Thaler}, \citenamefont {Hofman},
  \citenamefont {Xenias}, \citenamefont {Sims}, \citenamefont {Giannone},
  \citenamefont {Dustin}, \citenamefont {Wiggins},\ and\ \citenamefont
  {Sheetz}}]{Dobereiner2006}%
  \BibitemOpen
  \bibfield  {author} {\bibinfo {author} {\bibfnamefont {H.~G.}\ \bibnamefont
  {D{\"{o}}bereiner}}, \bibinfo {author} {\bibfnamefont {B.~J.}\ \bibnamefont
  {Dubin-Thaler}}, \bibinfo {author} {\bibfnamefont {J.~M.}\ \bibnamefont
  {Hofman}}, \bibinfo {author} {\bibfnamefont {H.~S.}\ \bibnamefont {Xenias}},
  \bibinfo {author} {\bibfnamefont {T.~N.}\ \bibnamefont {Sims}}, \bibinfo
  {author} {\bibfnamefont {G.}~\bibnamefont {Giannone}}, \bibinfo {author}
  {\bibfnamefont {M.~L.}\ \bibnamefont {Dustin}}, \bibinfo {author}
  {\bibfnamefont {C.~H.}\ \bibnamefont {Wiggins}}, \ and\ \bibinfo {author}
  {\bibfnamefont {M.~P.}\ \bibnamefont {Sheetz}},\ }\bibfield  {title}
  {\enquote {\bibinfo {title} {{Lateral membrane waves constitute a universal
  dynamic pattern of motile cells}},}\ }\href {\doibase
  10.1103/PhysRevLett.97.038102} {\bibfield  {journal} {\bibinfo  {journal}
  {Physical Review Letters}\ }\textbf {\bibinfo {volume} {97}},\ \bibinfo
  {pages} {10--13} (\bibinfo {year} {2006})}\BibitemShut {NoStop}%
\bibitem [{\citenamefont {Alonso}, \citenamefont {Bar},\ and\ \citenamefont
  {Echebarria}(2016)}]{Alonso2016}%
  \BibitemOpen
  \bibfield  {author} {\bibinfo {author} {\bibfnamefont {S.}~\bibnamefont
  {Alonso}}, \bibinfo {author} {\bibfnamefont {M.}~\bibnamefont {Bar}}, \ and\
  \bibinfo {author} {\bibfnamefont {B.}~\bibnamefont {Echebarria}},\ }\bibfield
   {title} {\enquote {\bibinfo {title} {{Nonlinear physics of electrical wave
  propagation in the heart: A review}},}\ }\href {\doibase
  10.1088/0034-4885/79/9/096601} {\bibfield  {journal} {\bibinfo  {journal}
  {Reports on Progress in Physics}\ }\textbf {\bibinfo {volume} {79}},\
  \bibinfo {pages} {96601} (\bibinfo {year} {2016})}\BibitemShut {NoStop}%
\bibitem [{\citenamefont {Chun}(2013)}]{Chun2013}%
  \BibitemOpen
  \bibfield  {author} {\bibinfo {author} {\bibfnamefont {S.}~\bibnamefont
  {Chun}},\ }\bibfield  {title} {\enquote {\bibinfo {title} {{Geometric
  analysis on the unidirectionality of the pulmonary veins for atrial
  reentry}},}\ }\href@noop {} {\  (\bibinfo {year} {2013})},\ \Eprint
  {http://arxiv.org/abs/1310.2466v1} {arXiv:1310.2466v1} \BibitemShut {NoStop}%
\bibitem [{\citenamefont {Neic}\ \emph {et~al.}(2017)\citenamefont {Neic},
  \citenamefont {Campos}, \citenamefont {Prassl}, \citenamefont {Niederer},
  \citenamefont {Bishop}, \citenamefont {Vigmond},\ and\ \citenamefont
  {Plank}}]{Neic2017}%
  \BibitemOpen
  \bibfield  {author} {\bibinfo {author} {\bibfnamefont {A.}~\bibnamefont
  {Neic}}, \bibinfo {author} {\bibfnamefont {F.~O.}\ \bibnamefont {Campos}},
  \bibinfo {author} {\bibfnamefont {A.~J.}\ \bibnamefont {Prassl}}, \bibinfo
  {author} {\bibfnamefont {S.~A.}\ \bibnamefont {Niederer}}, \bibinfo {author}
  {\bibfnamefont {M.~J.}\ \bibnamefont {Bishop}}, \bibinfo {author}
  {\bibfnamefont {E.~J.}\ \bibnamefont {Vigmond}}, \ and\ \bibinfo {author}
  {\bibfnamefont {G.}~\bibnamefont {Plank}},\ }\bibfield  {title} {\enquote
  {\bibinfo {title} {{Efficient computation of electrograms and ECGs in human
  whole heart simulations using a reaction-eikonal model}},}\ }\href {\doibase
  10.1016/j.jcp.2017.06.020} {\bibfield  {journal} {\bibinfo  {journal}
  {Journal of Computational Physics}\ }\textbf {\bibinfo {volume} {346}},\
  \bibinfo {pages} {191--211} (\bibinfo {year} {2017})}\BibitemShut {NoStop}%
\bibitem [{\citenamefont {Lubenov}\ and\ \citenamefont
  {Siapas}(2009)}]{Lubenov2009}%
  \BibitemOpen
  \bibfield  {author} {\bibinfo {author} {\bibfnamefont {E.~V.}\ \bibnamefont
  {Lubenov}}\ and\ \bibinfo {author} {\bibfnamefont {A.~G.}\ \bibnamefont
  {Siapas}},\ }\bibfield  {title} {\enquote {\bibinfo {title} {{Hippocampal
  theta oscillations are travelling waves}},}\ }\href {\doibase
  10.1038/nature08010} {\bibfield  {journal} {\bibinfo  {journal} {Nature}\
  }\textbf {\bibinfo {volume} {459}},\ \bibinfo {pages} {534--539} (\bibinfo
  {year} {2009})}\BibitemShut {NoStop}%
\bibitem [{\citenamefont {Agarwal}\ \emph {et~al.}(2014)\citenamefont
  {Agarwal}, \citenamefont {Stevenson}, \citenamefont {Berenyi}, \citenamefont
  {Mizuseki}, \citenamefont {Buzsaki},\ and\ \citenamefont
  {Sommer}}]{Agarwal2014}%
  \BibitemOpen
  \bibfield  {author} {\bibinfo {author} {\bibfnamefont {G.}~\bibnamefont
  {Agarwal}}, \bibinfo {author} {\bibfnamefont {I.~H.}\ \bibnamefont
  {Stevenson}}, \bibinfo {author} {\bibfnamefont {A.}~\bibnamefont {Berenyi}},
  \bibinfo {author} {\bibfnamefont {K.}~\bibnamefont {Mizuseki}}, \bibinfo
  {author} {\bibfnamefont {G.}~\bibnamefont {Buzsaki}}, \ and\ \bibinfo
  {author} {\bibfnamefont {F.~T.}\ \bibnamefont {Sommer}},\ }\bibfield  {title}
  {\enquote {\bibinfo {title} {{Spatially Distributed Local Fields in the
  Hippocampus Encode Rat Position}},}\ }\href {\doibase
  10.1126/science.1250444} {\bibfield  {journal} {\bibinfo  {journal}
  {Science}\ }\textbf {\bibinfo {volume} {344}},\ \bibinfo {pages} {626--630}
  (\bibinfo {year} {2014})}\BibitemShut {NoStop}%
\bibitem [{\citenamefont {Rubino}, \citenamefont {Robbins},\ and\ \citenamefont
  {Hatsopoulos}(2006)}]{Rubino2006}%
  \BibitemOpen
  \bibfield  {author} {\bibinfo {author} {\bibfnamefont {D.}~\bibnamefont
  {Rubino}}, \bibinfo {author} {\bibfnamefont {K.~A.}\ \bibnamefont {Robbins}},
  \ and\ \bibinfo {author} {\bibfnamefont {N.~G.}\ \bibnamefont
  {Hatsopoulos}},\ }\bibfield  {title} {\enquote {\bibinfo {title}
  {{Propagating waves mediate information transfer in the motor cortex}},}\
  }\href {\doibase 10.1038/nn1802} {\bibfield  {journal} {\bibinfo  {journal}
  {Nature Neuroscience}\ }\textbf {\bibinfo {volume} {9}},\ \bibinfo {pages}
  {1549--1557} (\bibinfo {year} {2006})}\BibitemShut {NoStop}%
\bibitem [{\citenamefont {Zanos}\ \emph {et~al.}(2015)\citenamefont {Zanos},
  \citenamefont {Mineault}, \citenamefont {Nasiotis}, \citenamefont {Guitton},\
  and\ \citenamefont {Pack}}]{Zanos2015}%
  \BibitemOpen
  \bibfield  {author} {\bibinfo {author} {\bibfnamefont {T.~P.}\ \bibnamefont
  {Zanos}}, \bibinfo {author} {\bibfnamefont {P.~J.}\ \bibnamefont {Mineault}},
  \bibinfo {author} {\bibfnamefont {K.~T.}\ \bibnamefont {Nasiotis}}, \bibinfo
  {author} {\bibfnamefont {D.}~\bibnamefont {Guitton}}, \ and\ \bibinfo
  {author} {\bibfnamefont {C.~C.}\ \bibnamefont {Pack}},\ }\bibfield  {title}
  {\enquote {\bibinfo {title} {{A Sensorimotor Role for Traveling Waves in
  Primate Visual Cortex}},}\ }\href {\doibase 10.1016/j.neuron.2014.12.043}
  {\bibfield  {journal} {\bibinfo  {journal} {Neuron}\ }\textbf {\bibinfo
  {volume} {85}},\ \bibinfo {pages} {615--627} (\bibinfo {year}
  {2015})}\BibitemShut {NoStop}%
\bibitem [{\citenamefont {Martinet}\ \emph {et~al.}(2017)\citenamefont
  {Martinet}, \citenamefont {Fiddyment}, \citenamefont {Madsen}, \citenamefont
  {Eskandar}, \citenamefont {Truccolo}, \citenamefont {Eden}, \citenamefont
  {Cash},\ and\ \citenamefont {Kramer}}]{Martinet2017}%
  \BibitemOpen
  \bibfield  {author} {\bibinfo {author} {\bibfnamefont {L.~E.}\ \bibnamefont
  {Martinet}}, \bibinfo {author} {\bibfnamefont {G.}~\bibnamefont {Fiddyment}},
  \bibinfo {author} {\bibfnamefont {J.~R.}\ \bibnamefont {Madsen}}, \bibinfo
  {author} {\bibfnamefont {E.~N.}\ \bibnamefont {Eskandar}}, \bibinfo {author}
  {\bibfnamefont {W.}~\bibnamefont {Truccolo}}, \bibinfo {author}
  {\bibfnamefont {U.~T.}\ \bibnamefont {Eden}}, \bibinfo {author}
  {\bibfnamefont {S.~S.}\ \bibnamefont {Cash}}, \ and\ \bibinfo {author}
  {\bibfnamefont {M.~A.}\ \bibnamefont {Kramer}},\ }\bibfield  {title}
  {\enquote {\bibinfo {title} {{Human seizures couple across spatial scales
  through travelling wave dynamics}},}\ }\href {\doibase 10.1038/ncomms14896}
  {\bibfield  {journal} {\bibinfo  {journal} {Nature Communications}\ }\textbf
  {\bibinfo {volume} {8}},\ \bibinfo {pages} {1--13} (\bibinfo {year}
  {2017})}\BibitemShut {NoStop}%
\bibitem [{\citenamefont {Santos}\ \emph {et~al.}(2014)\citenamefont {Santos},
  \citenamefont {Sch{\"{o}}ll}, \citenamefont {S{\'{a}}nchez-Porras},
  \citenamefont {Dahlem}, \citenamefont {Silos}, \citenamefont {Unterberg},
  \citenamefont {Dickhaus},\ and\ \citenamefont {Sakowitz}}]{Santos2014}%
  \BibitemOpen
  \bibfield  {author} {\bibinfo {author} {\bibfnamefont {E.}~\bibnamefont
  {Santos}}, \bibinfo {author} {\bibfnamefont {M.}~\bibnamefont
  {Sch{\"{o}}ll}}, \bibinfo {author} {\bibfnamefont {R.}~\bibnamefont
  {S{\'{a}}nchez-Porras}}, \bibinfo {author} {\bibfnamefont {M.~a.}\
  \bibnamefont {Dahlem}}, \bibinfo {author} {\bibfnamefont {H.}~\bibnamefont
  {Silos}}, \bibinfo {author} {\bibfnamefont {A.}~\bibnamefont {Unterberg}},
  \bibinfo {author} {\bibfnamefont {H.}~\bibnamefont {Dickhaus}}, \ and\
  \bibinfo {author} {\bibfnamefont {O.~W.}\ \bibnamefont {Sakowitz}},\
  }\bibfield  {title} {\enquote {\bibinfo {title} {{Radial, spiral and
  reverberating waves of spreading depolarization occur in the gyrencephalic
  brain}},}\ }\href {\doibase 10.1016/j.neuroimage.2014.05.021} {\bibfield
  {journal} {\bibinfo  {journal} {NeuroImage}\ }\textbf {\bibinfo {volume}
  {99}},\ \bibinfo {pages} {244--255} (\bibinfo {year} {2014})}\BibitemShut
  {NoStop}%
\bibitem [{\citenamefont {Verkhlyutov}\ and\ \citenamefont
  {Balaev}(2018)}]{Verkhlyutov2018}%
  \BibitemOpen
  \bibfield  {author} {\bibinfo {author} {\bibfnamefont {V.~M.}\ \bibnamefont
  {Verkhlyutov}}\ and\ \bibinfo {author} {\bibfnamefont {V.~V.}\ \bibnamefont
  {Balaev}},\ }\bibfield  {title} {\enquote {\bibinfo {title} {{The method of
  modeling the human EEG by calculating radial traveling waves on the folded
  surface of the human cerebral cortex}},}\ }\href {\doibase 10.1101/242412}
  {\bibfield  {journal} {\bibinfo  {journal} {bioRxiv}\ ,\ \bibinfo {pages}
  {242412}} (\bibinfo {year} {2018})}\BibitemShut {NoStop}%
\bibitem [{\citenamefont {Roberts}\ \emph {et~al.}(2019)\citenamefont
  {Roberts}, \citenamefont {Gollo}, \citenamefont {Abeysuriya}, \citenamefont
  {Roberts}, \citenamefont {Mitchell}, \citenamefont {Woolrich},\ and\
  \citenamefont {Breakspear}}]{Roberts2019}%
  \BibitemOpen
  \bibfield  {author} {\bibinfo {author} {\bibfnamefont {J.~A.}\ \bibnamefont
  {Roberts}}, \bibinfo {author} {\bibfnamefont {L.~L.}\ \bibnamefont {Gollo}},
  \bibinfo {author} {\bibfnamefont {R.~G.}\ \bibnamefont {Abeysuriya}},
  \bibinfo {author} {\bibfnamefont {G.}~\bibnamefont {Roberts}}, \bibinfo
  {author} {\bibfnamefont {P.~B.}\ \bibnamefont {Mitchell}}, \bibinfo {author}
  {\bibfnamefont {M.~W.}\ \bibnamefont {Woolrich}}, \ and\ \bibinfo {author}
  {\bibfnamefont {M.}~\bibnamefont {Breakspear}},\ }\bibfield  {title}
  {\enquote {\bibinfo {title} {{Metastable brain waves}},}\ }\href {\doibase
  10.1038/s41467-019-08999-0} {\bibfield  {journal} {\bibinfo  {journal}
  {Nature Communications}\ }\textbf {\bibinfo {volume} {10}},\ \bibinfo {pages}
  {1--17} (\bibinfo {year} {2019})}\BibitemShut {NoStop}%
\bibitem [{\citenamefont {H{\"{o}}rning}\ and\ \citenamefont
  {Shibata}(2019)}]{Horning2019}%
  \BibitemOpen
  \bibfield  {author} {\bibinfo {author} {\bibfnamefont {M.}~\bibnamefont
  {H{\"{o}}rning}}\ and\ \bibinfo {author} {\bibfnamefont {T.}~\bibnamefont
  {Shibata}},\ }\bibfield  {title} {\enquote {\bibinfo {title}
  {{Three-Dimensional Cell Geometry Controls Excitable Membrane Signaling in
  Dictyostelium Cells}},}\ }\href {\doibase 10.1016/j.bpj.2018.12.012}
  {\bibfield  {journal} {\bibinfo  {journal} {Biophysical Journal}\ }\textbf
  {\bibinfo {volume} {116}},\ \bibinfo {pages} {372--382} (\bibinfo {year}
  {2019})}\BibitemShut {NoStop}%
\bibitem [{\citenamefont {Gerhardt}\ \emph
  {et~al.}(2014{\natexlab{b}})\citenamefont {Gerhardt}, \citenamefont {Ecke},
  \citenamefont {Walz}, \citenamefont {Stengl}, \citenamefont {Beta},\ and\
  \citenamefont {Gerisch}}]{Gerhardt2014a}%
  \BibitemOpen
  \bibfield  {author} {\bibinfo {author} {\bibfnamefont {M.}~\bibnamefont
  {Gerhardt}}, \bibinfo {author} {\bibfnamefont {M.}~\bibnamefont {Ecke}},
  \bibinfo {author} {\bibfnamefont {M.}~\bibnamefont {Walz}}, \bibinfo {author}
  {\bibfnamefont {A.}~\bibnamefont {Stengl}}, \bibinfo {author} {\bibfnamefont
  {C.}~\bibnamefont {Beta}}, \ and\ \bibinfo {author} {\bibfnamefont
  {G.}~\bibnamefont {Gerisch}},\ }\bibfield  {title} {\enquote {\bibinfo
  {title} {{Actin and PIP3 waves in giant cells reveal the inherent length
  scale of an excited state.}}}\ }\href {\doibase 10.1242/jcs.156000}
  {\bibfield  {journal} {\bibinfo  {journal} {Journal of cell science}\
  }\textbf {\bibinfo {volume} {127}},\ \bibinfo {pages} {4507--17} (\bibinfo
  {year} {2014}{\natexlab{b}})}\BibitemShut {NoStop}%
\bibitem [{\citenamefont {Heitmann}, \citenamefont {Boonstra},\ and\
  \citenamefont {Breakspear}(2013)}]{Heitmann2013}%
  \BibitemOpen
  \bibfield  {author} {\bibinfo {author} {\bibfnamefont {S.}~\bibnamefont
  {Heitmann}}, \bibinfo {author} {\bibfnamefont {T.}~\bibnamefont {Boonstra}},
  \ and\ \bibinfo {author} {\bibfnamefont {M.}~\bibnamefont {Breakspear}},\
  }\bibfield  {title} {\enquote {\bibinfo {title} {{A Dendritic Mechanism for
  Decoding Traveling Waves: Principles and Applications to Motor Cortex}},}\
  }\href {\doibase 10.1371/journal.pcbi.1003260} {\bibfield  {journal}
  {\bibinfo  {journal} {PLoS Computational Biology}\ }\textbf {\bibinfo
  {volume} {9}} (\bibinfo {year} {2013}),\
  10.1371/journal.pcbi.1003260}\BibitemShut {NoStop}%
\bibitem [{\citenamefont {Mimura}\ and\ \citenamefont
  {Nagayama}(1997)}]{Mimura1997a}%
  \BibitemOpen
  \bibfield  {author} {\bibinfo {author} {\bibfnamefont {M.}~\bibnamefont
  {Mimura}}\ and\ \bibinfo {author} {\bibfnamefont {M.}~\bibnamefont
  {Nagayama}},\ }\bibfield  {title} {\enquote {\bibinfo {title}
  {{Nonannihilation dynamics in an exothermic reaction-diffusion system with
  mono-stable excitability.}}}\ }\href {\doibase 10.1063/1.166282} {\bibfield
  {journal} {\bibinfo  {journal} {Chaos (Woodbury, N.Y.)}\ }\textbf {\bibinfo
  {volume} {7}},\ \bibinfo {pages} {817--826} (\bibinfo {year}
  {1997})}\BibitemShut {NoStop}%
\bibitem [{\citenamefont {Ei}, \citenamefont {Mimura},\ and\ \citenamefont
  {Nagayama}(2002)}]{Ei2002a}%
  \BibitemOpen
  \bibfield  {author} {\bibinfo {author} {\bibfnamefont {S.}~\bibnamefont
  {Ei}}, \bibinfo {author} {\bibfnamefont {M.}~\bibnamefont {Mimura}}, \ and\
  \bibinfo {author} {\bibfnamefont {M.}~\bibnamefont {Nagayama}},\ }\bibfield
  {title} {\enquote {\bibinfo {title} {{Pulse – pulse interaction in reaction
  – diffusion systems}},}\ }\href@noop {} {\bibfield  {journal} {\bibinfo
  {journal} {Physica D}\ }\textbf {\bibinfo {volume} {165}},\ \bibinfo {pages}
  {176--198} (\bibinfo {year} {2002})}\BibitemShut {NoStop}%
\bibitem [{\citenamefont {Jones}(1984)}]{Jones1984}%
  \BibitemOpen
  \bibfield  {author} {\bibinfo {author} {\bibfnamefont {C.~K. R.~T.}\
  \bibnamefont {Jones}},\ }\bibfield  {title} {\enquote {\bibinfo {title}
  {{Stability of the travelling wave solution of the FitzHugh-Nagumo
  system}},}\ }\href {\doibase 10.1090/S0002-9947-1984-0760971-6} {\bibfield
  {journal} {\bibinfo  {journal} {Transactions of the American Mathematical
  Society}\ }\textbf {\bibinfo {volume} {286}},\ \bibinfo {pages} {431--431}
  (\bibinfo {year} {1984})}\BibitemShut {NoStop}%
\bibitem [{\citenamefont {Tsujikawa}\ \emph {et~al.}(1989)\citenamefont
  {Tsujikawa}, \citenamefont {Nagai}, \citenamefont {Mimura}, \citenamefont
  {Kobayashi},\ and\ \citenamefont {Ikeda}}]{Tsujikawa1989a}%
  \BibitemOpen
  \bibfield  {author} {\bibinfo {author} {\bibfnamefont {T.}~\bibnamefont
  {Tsujikawa}}, \bibinfo {author} {\bibfnamefont {T.}~\bibnamefont {Nagai}},
  \bibinfo {author} {\bibfnamefont {M.}~\bibnamefont {Mimura}}, \bibinfo
  {author} {\bibfnamefont {R.}~\bibnamefont {Kobayashi}}, \ and\ \bibinfo
  {author} {\bibfnamefont {H.}~\bibnamefont {Ikeda}},\ }\bibfield  {title}
  {\enquote {\bibinfo {title} {{Stability properties of traveling pulse
  solutions of the higher dimensional FitzHugh-Nagumo equations}},}\ }\href
  {\doibase 10.1007/BF03167885} {\bibfield  {journal} {\bibinfo  {journal}
  {Japan Journal of Applied Mathematics}\ }\textbf {\bibinfo {volume} {6}},\
  \bibinfo {pages} {341--366} (\bibinfo {year} {1989})}\BibitemShut {NoStop}%
\bibitem [{\citenamefont {Biton}\ \emph {et~al.}(2009)\citenamefont {Biton},
  \citenamefont {Rabinovitch}, \citenamefont {Aviram},\ and\ \citenamefont
  {Braunstein}}]{Biton2009a}%
  \BibitemOpen
  \bibfield  {author} {\bibinfo {author} {\bibfnamefont {Y.}~\bibnamefont
  {Biton}}, \bibinfo {author} {\bibfnamefont {A.}~\bibnamefont {Rabinovitch}},
  \bibinfo {author} {\bibfnamefont {I.}~\bibnamefont {Aviram}}, \ and\ \bibinfo
  {author} {\bibfnamefont {D.}~\bibnamefont {Braunstein}},\ }\bibfield  {title}
  {\enquote {\bibinfo {title} {{Two mechanisms of spiral-pair-source creation
  in excitable media}},}\ }\href {\doibase 10.1016/j.physleta.2009.03.030}
  {\bibfield  {journal} {\bibinfo  {journal} {Physics Letters, Section A:
  General, Atomic and Solid State Physics}\ }\textbf {\bibinfo {volume}
  {373}},\ \bibinfo {pages} {1762--1767} (\bibinfo {year} {2009})}\BibitemShut
  {NoStop}%
\bibitem [{\citenamefont {Davydov}, \citenamefont {Morozov},\ and\
  \citenamefont {Davydov}(2003)}]{Davydov2003}%
  \BibitemOpen
  \bibfield  {author} {\bibinfo {author} {\bibfnamefont {V.~A.}\ \bibnamefont
  {Davydov}}, \bibinfo {author} {\bibfnamefont {V.~G.}\ \bibnamefont
  {Morozov}}, \ and\ \bibinfo {author} {\bibfnamefont {N.~V.}\ \bibnamefont
  {Davydov}},\ }\bibfield  {title} {\enquote {\bibinfo {title} {{Critical
  properties of autowaves propagating on deformed cylindrical surfaces}},}\
  }\href {\doibase 10.1016/S0375-9601(02)01726-7} {\bibfield  {journal}
  {\bibinfo  {journal} {Physics Letters, Section A: General, Atomic and Solid
  State Physics}\ }\textbf {\bibinfo {volume} {307}},\ \bibinfo {pages}
  {265--268} (\bibinfo {year} {2003})}\BibitemShut {NoStop}%
\bibitem [{\citenamefont {Kneer}, \citenamefont {Sch{\"{o}}ll},\ and\
  \citenamefont {Dahlem}(2014)}]{Kneer2014}%
  \BibitemOpen
  \bibfield  {author} {\bibinfo {author} {\bibfnamefont {F.}~\bibnamefont
  {Kneer}}, \bibinfo {author} {\bibfnamefont {E.}~\bibnamefont {Sch{\"{o}}ll}},
  \ and\ \bibinfo {author} {\bibfnamefont {M.~A.}\ \bibnamefont {Dahlem}},\
  }\bibfield  {title} {\enquote {\bibinfo {title} {{Nucleation of
  reaction-diffusion waves on curved surfaces}},}\ }\href@noop {} {\bibfield
  {journal} {\bibinfo  {journal} {New Journal of Physics}\ }\textbf {\bibinfo
  {volume} {16}} (\bibinfo {year} {2014})}\BibitemShut {NoStop}%
\bibitem [{\citenamefont {Macdonald}, \citenamefont {Merriman},\ and\
  \citenamefont {Ruuth}(2013)}]{Macdonald2013}%
  \BibitemOpen
  \bibfield  {author} {\bibinfo {author} {\bibfnamefont {C.~B.}\ \bibnamefont
  {Macdonald}}, \bibinfo {author} {\bibfnamefont {B.}~\bibnamefont {Merriman}},
  \ and\ \bibinfo {author} {\bibfnamefont {S.~J.}\ \bibnamefont {Ruuth}},\
  }\bibfield  {title} {\enquote {\bibinfo {title} {{Simple computation of
  reaction–diffusion processes on point clouds}},}\ }\href {\doibase
  10.1073/pnas.1221408110} {\bibfield  {journal} {\bibinfo  {journal}
  {Proceedings of the National Academy of Sciences}\ }\textbf {\bibinfo
  {volume} {110}},\ \bibinfo {pages} {9209--9214} (\bibinfo {year}
  {2013})}\BibitemShut {NoStop}%
\bibitem [{\citenamefont {Nagumo}, \citenamefont {Arimoto},\ and\ \citenamefont
  {Yoshizawa}(1962)}]{Nagumo1962}%
  \BibitemOpen
  \bibfield  {author} {\bibinfo {author} {\bibfnamefont {J.}~\bibnamefont
  {Nagumo}}, \bibinfo {author} {\bibfnamefont {S.}~\bibnamefont {Arimoto}}, \
  and\ \bibinfo {author} {\bibfnamefont {S.}~\bibnamefont {Yoshizawa}},\
  }\bibfield  {title} {\enquote {\bibinfo {title} {{An Active Pulse
  Transmission Line Simulating Nerve Axon}},}\ }\href {\doibase
  10.1109/JRPROC.1962.288235} {\bibfield  {journal} {\bibinfo  {journal}
  {Proceedings of the IRE}\ }\textbf {\bibinfo {volume} {50}},\ \bibinfo
  {pages} {2061--2070} (\bibinfo {year} {1962})}\BibitemShut {NoStop}%
\bibitem [{\citenamefont {FitzHugh}(1961)}]{FitzHugh1961}%
  \BibitemOpen
  \bibfield  {author} {\bibinfo {author} {\bibfnamefont {R.}~\bibnamefont
  {FitzHugh}},\ }\bibfield  {title} {\enquote {\bibinfo {title} {{Impulses and
  Physiological States in Theoretical Models of Nerve Membrane}},}\ }\href
  {\doibase 10.1016/S0006-3495(61)86902-6} {\bibfield  {journal} {\bibinfo
  {journal} {Biophysical Journal}\ }\textbf {\bibinfo {volume} {1}},\ \bibinfo
  {pages} {445--466} (\bibinfo {year} {1961})}\BibitemShut {NoStop}%
\bibitem [{\citenamefont {Kazhdan}, \citenamefont {Bolitho},\ and\
  \citenamefont {Hoppe}(2006)}]{Kazhdan2006}%
  \BibitemOpen
  \bibfield  {author} {\bibinfo {author} {\bibfnamefont {M.}~\bibnamefont
  {Kazhdan}}, \bibinfo {author} {\bibfnamefont {M.}~\bibnamefont {Bolitho}}, \
  and\ \bibinfo {author} {\bibfnamefont {H.}~\bibnamefont {Hoppe}},\ }\bibfield
   {title} {\enquote {\bibinfo {title} {{Poisson Surface Reconstruction}},}\
  }\href {\doibase 10.1145/1364901.1364904} {\bibfield  {journal} {\bibinfo
  {journal} {Proceedings of the Symposium on Geometry Processing}\ ,\ \bibinfo
  {pages} {61--70}} (\bibinfo {year} {2006})}\BibitemShut {NoStop}%
\bibitem [{\citenamefont {Ferziger}\ and\ \citenamefont
  {Peri{\'{c}}}(2002)}]{Ferziger2002}%
  \BibitemOpen
  \bibfield  {author} {\bibinfo {author} {\bibfnamefont {J.~H.}\ \bibnamefont
  {Ferziger}}\ and\ \bibinfo {author} {\bibfnamefont {M.}~\bibnamefont
  {Peri{\'{c}}}},\ }\href {http://link.springer.com/10.1007/978-3-642-56026-2}
  {\emph {\bibinfo {title} {{Computational Methods for Fluid Dynamics}}}}\
  (\bibinfo  {publisher} {Springer Science {\&} Business Media},\ \bibinfo
  {year} {2002})\BibitemShut {NoStop}%
\bibitem [{\citenamefont {de~Padua}, \citenamefont {Parisio-Filho},\ and\
  \citenamefont {Moraes}(1998)}]{DePadua1998}%
  \BibitemOpen
  \bibfield  {author} {\bibinfo {author} {\bibfnamefont {a.}~\bibnamefont
  {de~Padua}}, \bibinfo {author} {\bibfnamefont {F.}~\bibnamefont
  {Parisio-Filho}}, \ and\ \bibinfo {author} {\bibfnamefont {F.}~\bibnamefont
  {Moraes}},\ }\bibfield  {title} {\enquote {\bibinfo {title} {{Geodesics
  around line defects in elastic solids}},}\ }\href {\doibase
  10.1016/S0375-9601(97)00871-2} {\bibfield  {journal} {\bibinfo  {journal}
  {Physics Letters A}\ }\textbf {\bibinfo {volume} {238}},\ \bibinfo {pages}
  {153--158} (\bibinfo {year} {1998})}\BibitemShut {NoStop}%
\bibitem [{\citenamefont {Butcher}(1964)}]{Butcher1964}%
  \BibitemOpen
  \bibfield  {author} {\bibinfo {author} {\bibfnamefont {J.~C.}\ \bibnamefont
  {Butcher}},\ }\bibfield  {title} {\enquote {\bibinfo {title} {{Implicit
  Runge-Kutta Processes}},}\ }\href {\doibase 10.2307/2003405} {\bibfield
  {journal} {\bibinfo  {journal} {Mathematics of Computation}\ }\textbf
  {\bibinfo {volume} {18}},\ \bibinfo {pages} {50--64} (\bibinfo {year}
  {1964})}\BibitemShut {NoStop}%
\bibitem [{\citenamefont {Martin}\ \emph {et~al.}(2018)\citenamefont {Martin},
  \citenamefont {Chappell}, \citenamefont {Chuzhanova},\ and\ \citenamefont
  {Crofts}}]{Martin2018}%
  \BibitemOpen
  \bibfield  {author} {\bibinfo {author} {\bibfnamefont {R.}~\bibnamefont
  {Martin}}, \bibinfo {author} {\bibfnamefont {D.~J.}\ \bibnamefont
  {Chappell}}, \bibinfo {author} {\bibfnamefont {N.}~\bibnamefont
  {Chuzhanova}}, \ and\ \bibinfo {author} {\bibfnamefont {J.~J.}\ \bibnamefont
  {Crofts}},\ }\bibfield  {title} {\enquote {\bibinfo {title} {{A numerical
  simulation of neural fields on curved geometries}},}\ }\href {\doibase
  10.1007/s10827-018-0697-5} {\bibfield  {journal} {\bibinfo  {journal}
  {Journal of Computational Neuroscience}\ }\textbf {\bibinfo {volume} {45}},\
  \bibinfo {pages} {133--145} (\bibinfo {year} {2018})}\BibitemShut {NoStop}%
\end{thebibliography}
\end{document}